# Effect of pH, surface charge and counter-ions on the Adsorption of Sodium Dodecyl Sulfate to the Sapphire/Solution Interface

Ningning Li, Robert K. Thomas*

Physical and Theoretical Chemistry Laboratory, Oxford University, South Parks Road, Oxford, OX1 3QZ, UK.

Adrian R. Rennie*

Department of Physics and Astronomy, Uppsala University, Box 516, SE-75120 Uppsala, Sweden.

* Please address correspondence to these authors: Adrian R. Rennie, e-mail: Adrian.Rennie@fysik.uu.se or Robert K. Thomas, e-mail: robert.thomas@chem.ox.ac.uk

**Abstract**

The role of ionic interactions between sodium dodecyl sulfate, SDS, and sapphire surfaces have been studied using specular neutron reflection to determine the structure and composition of adsorbed surfactant layers. Increasing the pH of the solution from 3 to 9 reduces the adsorption by reversing the charge of the alumina. This occurs at lower pH for the R-plane ($1\bar{1}02$) than the C-plane (0 0 0 1), corresponding to the different points of zero charge. The largest surface excess is about 6.5 μmol m$^{-2}$, the thickness of the adsorbed layer is about 24 Å and it contains roughly 20% water. The hydrocarbon tails of the surfactant molecules clearly interpenetrate rather than form an ordered bilayer. The structure is similar in either pure water or in 0.1 M NaCl when the surfactant is at the respective critical micelle concentration. Different structures were seen with lithium and cesium dodecyl sulfate. The CsDS forms dense layers with little or no hydration and a surface excess of about 10.5 μmol m$^{-2}$. The metal cation strongly influences the hydration of the adsorbed surfactant. An overall picture of 'flattened micelles' for the structure of the adsorbed layer is observed.

**Highlights**

- Increasing pH markedly reduces the adsorption of sodium dodecyl sulfate to sapphire but does not alter the thickness of the adsorbed layer
- Different crystal faces show markedly different adsorption that can be correlated with the surface charge
- The adsorbed layers of SDS are thinner than expected for a bilayer of extended chains and are highly hydrated
- Cesium dodecyl sulfate forms densely packed thin layers, even well below the critical micelle concentration

Table of Contents Graphic

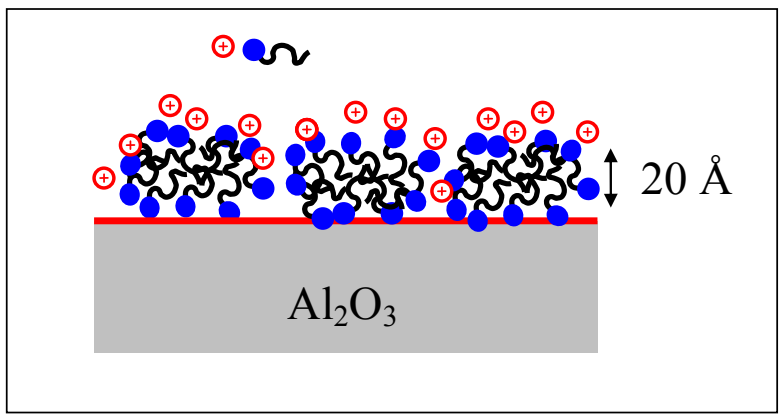

**Introduction**

Surfactants are widely used in industry, in domestic products such as detergents and for personal care, as well as in foods and for technology applications. In many cases, their role is to modify the properties of an interface. Surfactants are often used in aqueous solutions. Understanding adsorption processes from solution and the properties of adsorbed layers is thus of crucial importance in optimizing their use. A number of reviews [1,2] describe general features of surfactant adsorption to a broad range of hydrophilic interfaces. It is clear that a delicate balance of free energy of molecules bound at surfaces, dispersed as aggregates, or in solution may depend on fine details of molecular structure and interfacial chemistry. The wide range of behavior observed for different surfactants makes it important to relate structure and composition of adsorbed layers to physicochemical parameters of solutions and substrates.

Sodium dodecyl sulfate (SDS) is a relatively cheap anionic surfactant that is used widely in products such as shampoos, detergents and toothpaste. It is also used in mineral processing [3] for flotation of specific components and used to denature proteins in laboratory applications. There is wide general interest in binding of anionic surfactants to interfaces and specific interest in SDS at alumina surfaces. For example, there are possible applications of SDS to enhance corrosion resistance of aluminum oxide. [4] There is also recent interest in the application of SDS to act as a templating agent for growth of mesostructures such as alumina nanotubes. [5] This paper reports a study of the adsorption of SDS to the surfaces of crystalline alumina ($Al_2O_3$ or sapphire). In order to understand the influence of counter-ions, additional results with lithium and cesium salts of the dodecyl sulfate are also described.

**Background**

Solution properties, adsorption and micellisation of SDS have been widely studied. The critical micelle concentration (cmc) decreases as salt is added. [6] In pure water, it is about $8.2 \times 10^{-3}$ mol L$^{-1}$ and in the presence of 0.1 M NaCl, the cmc is $1.4 \times 10^{-3}$ mol L$^{-1}$. The Krafft boundary, at which crystallization occurs from a dispersion of micelles, for SDS in water is 16 °C. A number of papers describe the adsorption of SDS to colloidal alumina [7,8] and other substrates [9] using methods such as solution depletion or fluorescence spectroscopy. Although these experiments have identified that hydrophobic regions exist in adsorbed layers, indicative of some surfactant aggregation, they do not provide direct structural information. Atomic force microscopy [9] has indicated that there is some lateral structure with a periodicity of about 70 Å when SDS is adsorbed to self-assembled hydrocarbon monolayers on gold that are terminated with trimethyl ammonium ions. This has been taken to suggest that cylindrical micelles are adsorbed at that interface. The effects of changing temperature on the adsorption of SDS to $Al_2O_3$ were investigated explicitly by Sperline et al by attenuated total reflection infrared spectroscopy. [10] These studies extended to temperatures as low as 4 °C that are well below the Krafft boundary but the spectra from the interfacial surfactant layers were still characteristic of disordered structures, similar to micelles rather than crystals.

A significant challenge in the studies with SDS is hydrolysis of the sulfate group that can give rise to significant levels of impurities such as dodecanol and dodecanoic acid. It is important to purify the SDS immediately prior to use, for example by recrystallization. Studies of adsorption of SDS at the air/solution [11] and the solution/polystyrene [12] interfaces have specifically addressed the question of what happens to the impurities using deuterium labeled dodecanol and identified under what conditions dodecanol appears at the surface.

Adsorption of anionic surfactants to alumina is of particular interest as the charge at the sapphire water/interface can be controlled by adjusting the pH. Although sapphire is a pure oxide, the surface can be modified readily, and in the presence of water, there will be OH groups. At values of the pH below the isoelectric point, sapphire has a positive charge. The crystal of sapphire is trigonal and it is conventional to use four Miller indices to describe the crystal planes. The basal plane (0 0 0 1) is also known as the C-plane. Another crystal face that is widely used because of its facility for epitaxial growth of metals is the (1 $\bar{1}$ 0 2) or R-plane. The chemical composition and properties of the various faces are significantly different: there is a higher density of oxygen atoms accessible at the C-plane surface. The point of zero charge for the R-plane occurs at around pH 4.5 and increases to about pH 6 for the C-plane. [13]

The size, aggregation number and shape of SDS micelles has been reported in a number of studies. [14-19] These results give interesting ideas as to how SDS molecules pack in micelles and the consequences of screening charged interactions by adding salt. The earlier studies, e.g. [14] and [15], using small-angle neutron scattering were concerned primarily with the interactions between micelles. Bergström and Pedersen [17] present a careful analysis of the shape of micelles from scattering data and review the previous literature. As the surfactant concentration increases, the micelles grow into ellipsoidal or elongated shapes. Detailed experimental studies of SDS micelles have also prompted computer modeling [20,21] as efficient algorithms can be tested against reliable data. The influence of different metal cations on the behavior of alkyl sulfate surfactants has been investigated by a number of authors. Ridell et al have compared lithium, sodium and potassium ions with regard to the interaction of dodecyl sulfate with cellulose esters. [22] Micelles of lithium dodecyl sulfate [23] and cesium dodecyl sulfate [24] have been studied by small-angle neutron and X-ray scattering. There are also broader studies of a range of different counter-ions. [25-27] The influence of different metal ions with alkyl sulfate surfactants on adsorbed layers has been described for both aqueous [28] and polar, but non-aqueous solution interfaces. [29]

Specular reflection of neutrons provides a powerful tool to analyze the structure and composition of layers of molecules that are physisorbed or grafted to an interface. [30] It is particularly valuable as many solid materials, such as single crystals of sapphire, silicon or quartz are highly transparent to neutrons and so a neutron beam can be transmitted at grazing incidence through them in order to probe a solid/solution interface. Several reviews describe the application of this technique to study surfactant solutions at solid or vapor interfaces. [31,32] Scattering of neutrons from nuclei can depend on the particular isotope of an element. There is a large difference between hydrogen ($^1$H) and deuterium ($^2$H or D). This can be exploited to provide complementary data sets from chemically identical systems in order to avoid the ambiguity in data associated with unknown phase in scattering experiments. The changes of isotopic contrast or label can also be used to identify uniquely the composition of regions in a sample that consist of several components. The refractive index for neutrons of a material is calculated from atomic composition and density using tabulated data. [33,34] It is convenient to work with the scattering length density, $\rho$ that is calculated from the scattering lengths of the constituent atoms and the molecular volumes. Specular neutron reflectivity varies with wavelength, $\lambda$ and the grazing angle of the incidence of the beam, $\theta$. It depends on the density profile perpendicular to the interface. Reflectivity, R, is usually reported as a function of the momentum transfer vector, $Q = (4\pi/\lambda) \sin \theta$. The reflectivity for layered structures with known refractive index or scattering length density can be calculated using algorithms developed for optics with visible light [35] that are formulated in terms of composition and layer thickness. These calculations can also include a small interfacial roughness between adjacent layers. [36] The relationships between the 'optical' parameters and the practical quantities of interest such as surface excess are provided below in the section concerning interpretation of data.

Neutron reflection has provided information about adsorption of a wide range of surfactants. A wide-range of mixtures of surfactants and of surfactants and polymers that include SDS as one component has been studied. Directly relevant and comparable to this work are investigations of other ionic [37,38] and non-ionic surfactants [39] to sapphire.

**Experiments and Interpretation of Data**

*Reflection Experiments.* The neutron measurements were performed with the D17 reflectometer, Institut, Laue Langevin, Grenoble, France (ILL). [40] The measurements on the C-plane used two glancing angles of incidence, 0.5° and 3°, to produce reflectivity profiles over the momentum transfer range, $Q$ between 0.007 and 0.25Å$^{-1}$. On the R-plane, three incident angles 0.8°, 1.0°, 4.0° were used to give reflectivity over a similar $Q$ range. The range of $Q$ for which reflectivity could be measured depends on the magnitude of the signal above the background. In general, it was not feasible to record reflection much below $10^{-6}$. Translation scans to optimize the reflected intensity ensured that the sample surface was aligned with its center on the beam. The temperature of the substrates was maintained at 25±0.5 °C during all the neutron measurements by circulating water from a thermostat bath through the metal supports of the sample cells. The temperature was measured using a platinum resistance thermometer attached to the cell.

Data were reduced from the recorded counts by normalizing for the intensity of the incident beam and the transmission of the crystal. The background recorded on the two-dimensional detector in regions adjacent to the reflected beam was also subtracted using software provided at the ILL. The angle of reflection, used in calculating $Q$ for each measurement was determined from the position of the beam on the position sensitive detector making the experiment relatively insensitive to small misalignments of the sample angle.

*Materials.* SDS and lithium dodecyl sulfate (LiDS) were purchased from Fluka (BioUltra, ≥99.0%, GC). Cesium dodecyl sulfate (CsDS) was prepared by neutralizing dodecyl hydrogen sulfate with cesium hydroxide (Aldrich, 99%). The organic acid was prepared by ion exchanging SDS with Dowex H-form resin obtained from Sigma. The resin was washed three times with ultrapure water (UHQ) prior to the exchange reaction. All the samples were purified by repeated recrystallization from ethanol and washing with freshly distilled acetone. The surface tension was measured to verify that each of the samples were pure and that there was no minimum arising from impurities such as dodecanol. This data is shown in the supporting information as Figure S1.

The experiments used two different crystal faces of sapphire (R and C planes) as substrates: the 5 × 5 × 1.25 cm crystals were purchased from PI-KEM Ltd (Shropshire, UK). All crystals were cleaned in a mixture of 5:4:1 $H_2O/H_2SO_4/H_2O_2$ (dilute 'Piranha solution') [41] and then exposed to UV/Ozone for 30 min to remove contaminants [42] prior to the experiments. The freshly prepared crystals were kept in UHQ water in order to maintain clean surfaces until measurements started. All PTFE cells and glassware used were first cleaned with 5% Decon solution, and then rinsed extensively with UHQ water.

$D_2O$ (density 1100 g $L^{-1}$) was obtained from EURISO-TOP, CEA, Saclay. 50%$D_2O$ was prepared by mixing $D_2O$ and UHQ $H_2O$ with a volume ratio of 1:1. The solution pH was adjusted by diluting with small amounts of either HCl or NaOH.

*Modeling Neutron Reflection Data.* In the present experiments, adequate fits to the data were obtained with models that consisted of a single uniform layer with a small interfacial roughness. The scattering length density, $\rho$ and thus the refractive index for neutrons, is related to the volume fraction of two components, surfactant and water, in a layer:

$$\rho = \varphi_s \rho_s + \varphi_w \rho_w \qquad (1)$$

where $\varphi_s$ and $\varphi_w$ are the volume fractions of the surfactant and water respectively, and $\rho_s$ and $\rho_w$ their scattering length densities. The constraint $\varphi_s + \varphi_w = 1$ is used. If the scattering length density of the pure surfactant and water are known (see Table I), the determined value of $\rho$ can be used to deduce $\varphi_s$ as follows:

$$\varphi_s = (\rho - \rho_w) / (\rho_s - \rho_w) \qquad (2).$$

If the thickness is $t$, then the adsorbed amount of surfactant as volume per unit area is $t\,\varphi_s$ and the conventional surface excess, $\Gamma$, in moles per unit area is

$$\Gamma = t\,\varphi_s\,\rho_m / M \qquad (3)$$

where $\rho_m$ is the mass density of the surfactant (in g m$^{-3}$) and M is the relative molecular mass. The area per molecule is simply $1/(\Gamma\,N_A)$ where $N_A$ is Avogadro's constant. For hydrophilic surfaces where comparison with a bilayer is appropriate, it is convenient to consider the area, $A$, for two molecules as this would represent the lateral packing of individual molecules.

$$A = 2M / t\,\varphi_s\,\rho_m\,N_A \qquad (4).$$

In general, fits were constrained so that $t$ and $\Gamma$ for measurements with two contrasts of water were the same within experimental error.

**Results**

*Characterization of the substrate.* The clean sapphire surfaces were initially characterized by with data recorded with H$_2$O, 50%D$_2$O and D$_2$O. The data and fits are shown in Figure 1 and indicated that the surface was clean and had a roughness of 5 Å. This roughness of the substrate was included in the subsequent fits of data recorded for adsorbed layers.

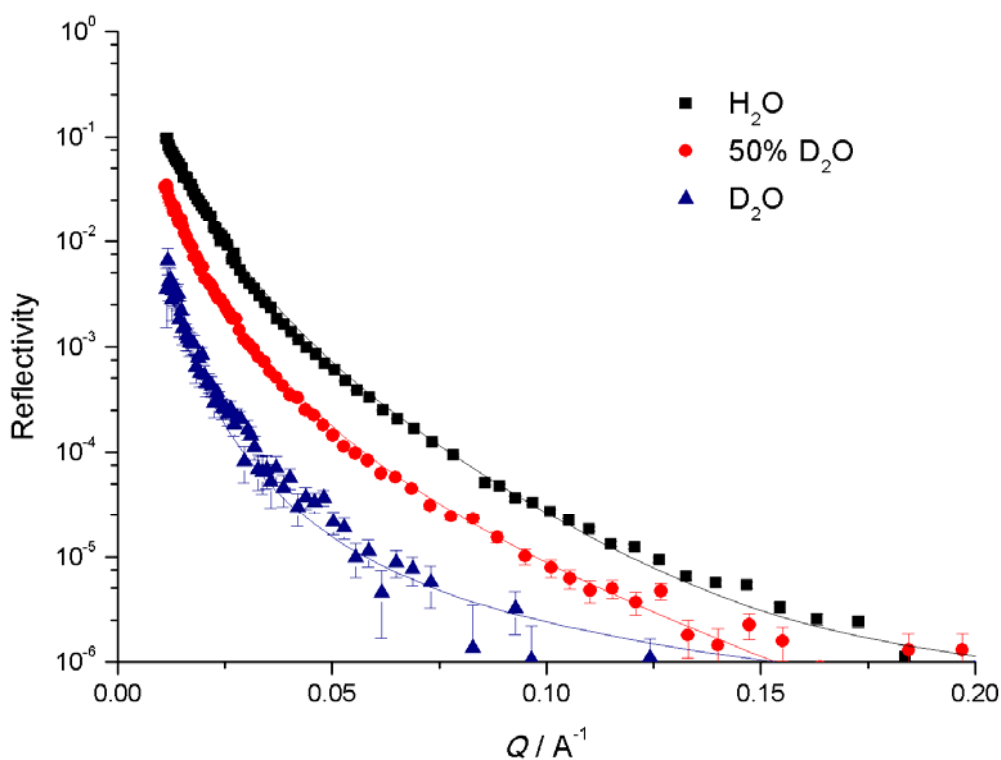

Figure 1. Neutron reflection data for the clean R-plane substrate with a model fit that includes 5 Å roughness at the water interface. Data is shown for $H_2O$, 50% $D_2O$ and $D_2O$. The calculated fits are shown as solid lines.

*Effect of pH on Adsorption.* The reflectivity data for the clean C-plane surface are, within experimental uncertainty, the same as that for the R plane. Both substrates were used for adsorption experiments using solutions of 1.4 mM SDS in 0.1 M NaCl. Data for solutions at different pH in $D_2O$ are shown for the R-plane and C-plane substrates in Figures 2 and 3 respectively. Some further data sets were measured with solutions in 50% $D_2O$ and these are shown in the supporting information as Figures S2 and S3. As the scattering length density of $D_2O$ is relatively close to that of sapphire and high while that of SDS is small, the reflectivity measured with a solution in $D_2O$ provides a simple indication of the amount of adsorbed surfactant. These could all be adequately modeled with a single uniform layer of surfactant and the parameters for the fits shown in the Figures are given in Tables II and III. The simple observation that the reflectivity decreases with increasing pH is apparent from the fitted model parameters in Table II that show the change in the surface excess. There is good consistency of the model fits as regards thickness and composition for the data measured with the two different solution contrasts.

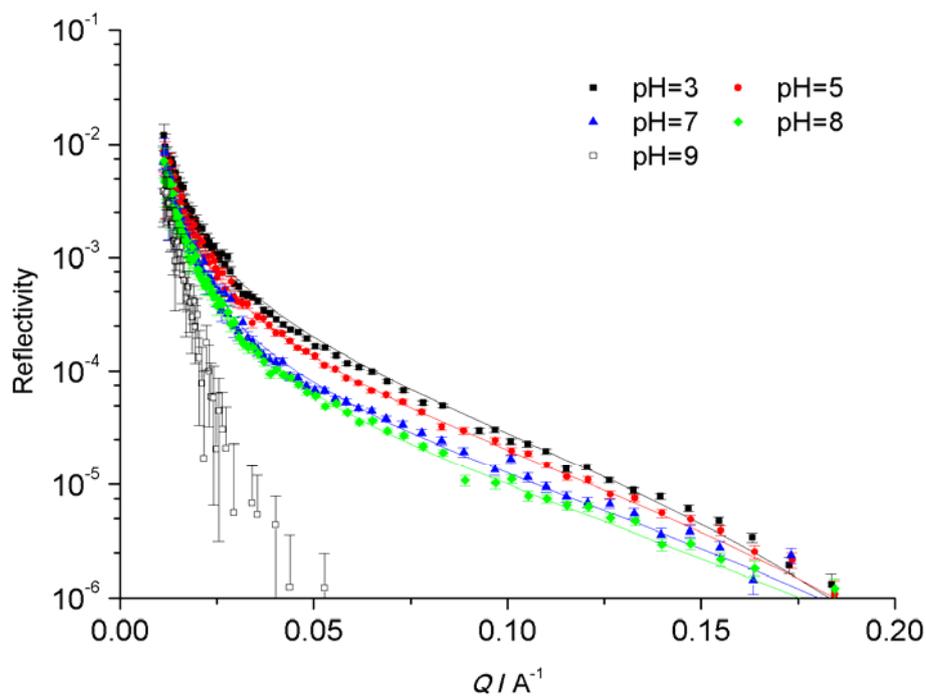

Figure 2. Neutron reflection data for 1.4 mM SDS in 0.1 M NaCl/$D_2O$ adsorbed on R-plane sapphire with uniform layer fits. The model parameters for the fits, shown as solid lines, are listed in Table II.

It may not be apparent that the shape of the reflection curves in Figure 2 is distinctive and provides a clear indication of the thickness of the adsorbed layer. The logarithmic scales permit a wide range of intensity to be visualized, readily facilitating comparison of the different data sets but an alternative plot of $RQ^4$ versus $Q$ shows clearly a strong peak. For this contrast condition, the position in $Q$ defines the thickness of the interfacial layer. This representation of the data for pH 3 in Figure 2 is in Figure 3.

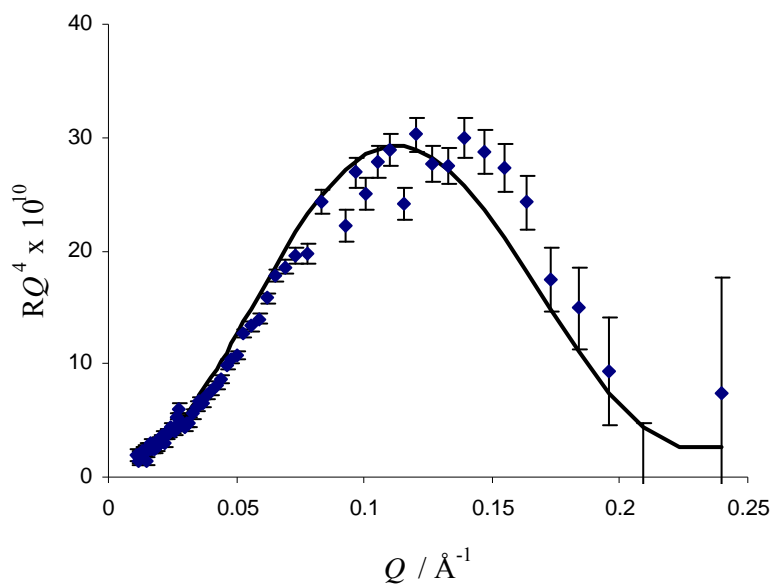

Figure 3. Alternative representation of the data and fit for pH 3 from Figure 2 for the solution with 1.4 mM SDS and 0.1 M NaCl and the R-plane substrate. The plot of $RQ^4$ versus $Q$ allows the position of the peak and the layer thickness to be observed directly.

Further measurements to explore the role of surface charge were made by studying the adsorption to a C-plane sapphire surface that has a higher point of zero charge. Data analogous to that in Figure 2 is shown in Figure 4 and the parameters for the fitted models are provided in Table III. It is clear that there are differences between the behavior of the surfactant at these interfaces. For the C-plane, data recorded at pH3 and pH5 are rather similar and changes are observed at higher pH.

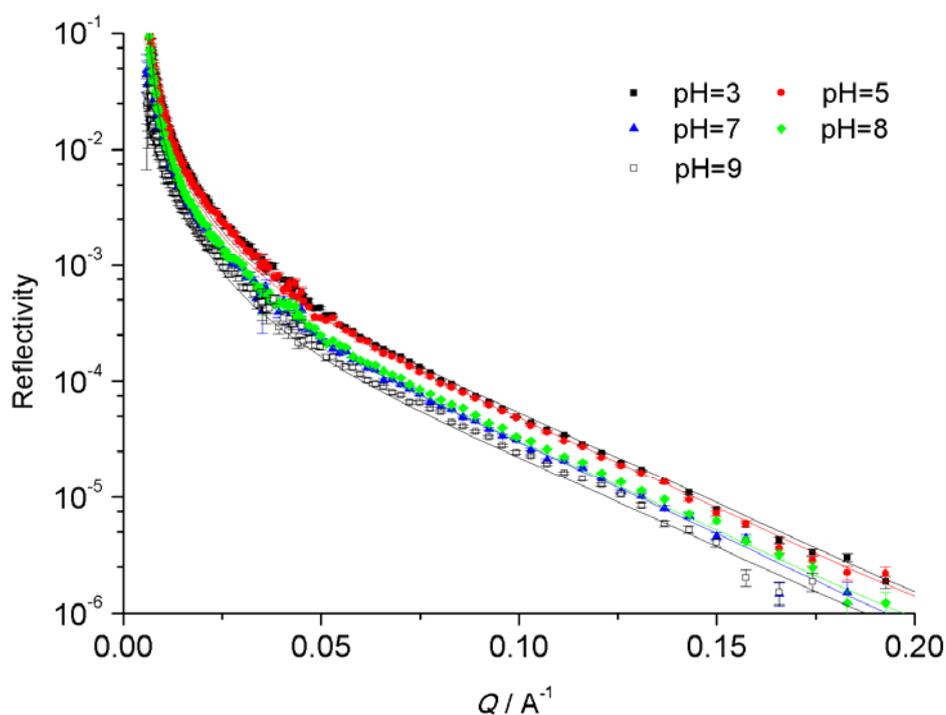

Figure 4. Neutron reflection data for 1.4 mM SDS in 0.1 M NaCl/$D_2O$ adsorbed on C-plane sapphire with uniform layer fits. The model parameters for the fits, shown as solid lines, are listed in Table III.

*Effect of Salt on Adsorption.* Some additional measurements were made with solutions without any added salt at neutral pH to investigate the role of extra salt in modifying the adsorption. The data for measurements of solutions with 2.4 and 8.2 mM SDS in $D_2O$ adsorbing on the C-plane sapphire are shown in Figure 5. Parameters for the fitted models are listed in Table IV. There is a clear increase of the amount of adsorbed surfactant as the concentration is increased from about 0.3 cmc to the cmc.

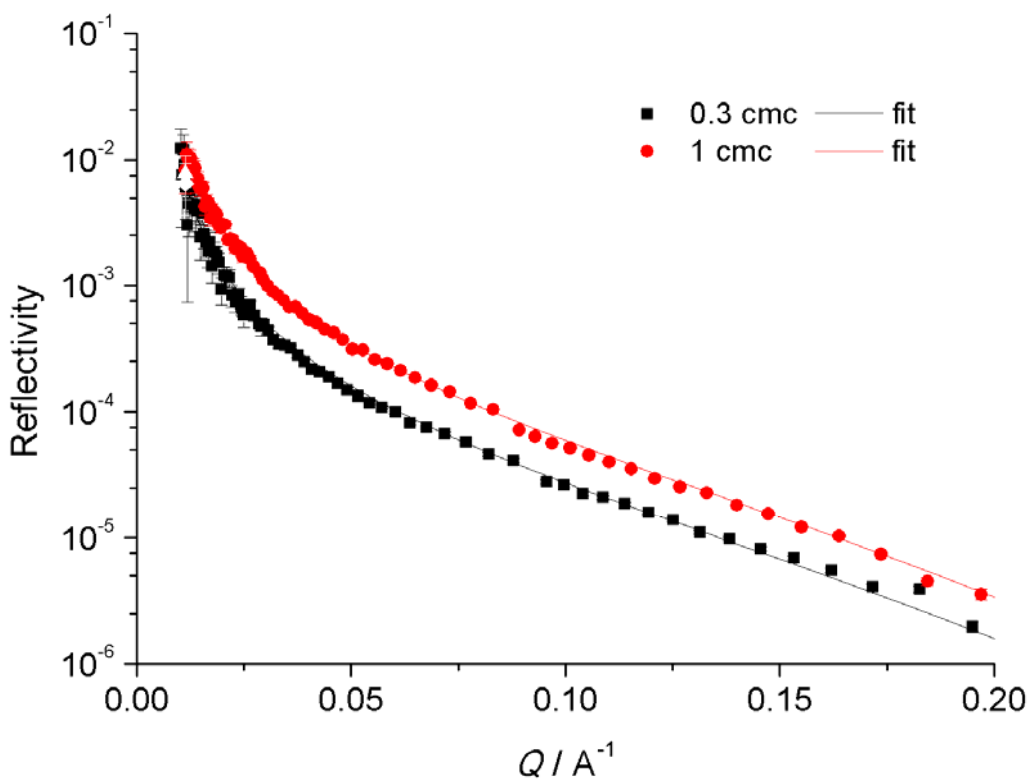

Figure 5. Neutron reflection data for SDS at concentrations of 2.4 mM (0.3 cmc) and 8.2 mM (1 cmc) in pure $D_2O$ adsorbed on C-plane sapphire with uniform layer fits. The model parameters for the fits are listed in Table IV.

In order to investigate the effect of different metal cations, additional data were measured for the samples of LiDS (at 25 °C) and CsDS. The measurements on LiDS were made at 25 °C but because the Krafft temperature of CsDS is 36 °C, [6] the measurements for this surfactant were made at 40 °C. These data were measured using the ADAM reflectometer at the ILL with a fixed wavelength of 4.4 Å and scans of angle. The data measured in $D_2O$ is shown in Figure 6. As with the other samples, measurements in 50% $D_2O$ were also made and used to find a unique single model to fit both data sets. The parameters for the models are also shown in Table IV.

The data recorded for solutions in 50% $D_2O$ is shown in Figures S4 and S5 in the supporting information.

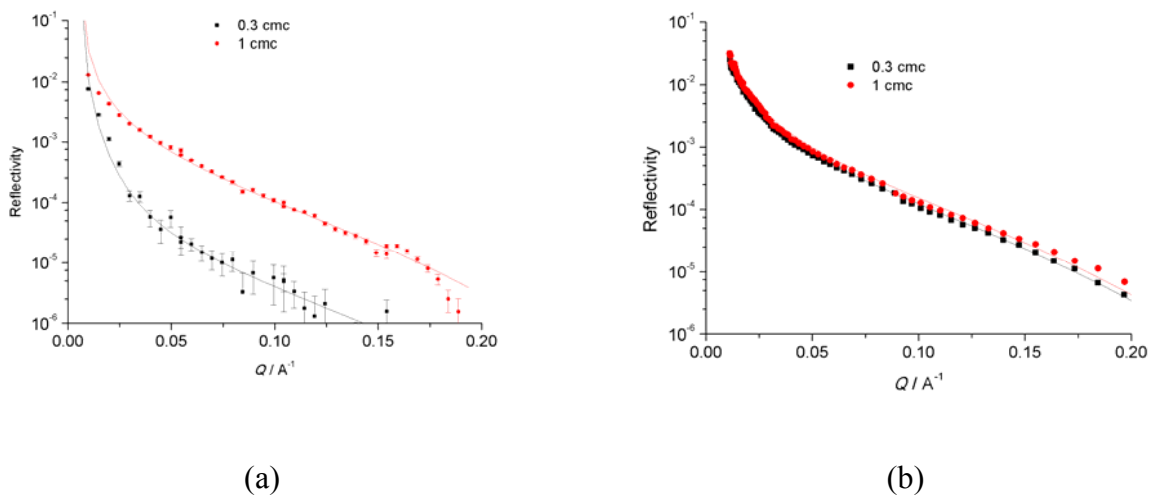

(a)　　　　　　　　　　　　　　　　　　(b)

Figure 6. Neutron reflectivity data for (a) LiDS and (b) CsDS in $D_2O$ measured at 25 and 40 °C respectively at concentrations of 0.3 cmc and 1 cmc.

**Discussion and Conclusions**

The data shown in Figure 2 display a clear trend as the pH changes from 3 to 9. The fitted parameters are shown in Figure 7. The isoelectric point of alumina depends on a number of factors. The various crystal faces have a different chemical composition and as mentioned above are expected to dissociate differently. There is no sharp break in the adsorption at the isoelectric point for either the R or the C crystal faces. There is rather a gradual decrease in the adsorbed amount as the pH increases. The C plane substrate shows a significantly higher adsorption at pH 6 as might be expected from the literature values of the respective points of zero charge. The maximum adsorption that was observed in these experiments, on the C-plane interface at pH 3, corresponds to about 6.5 μmol m$^{-2}$. This is similar to the amount observed on colloidal alumina. [43]

According to Tanford, [44] the maximum length, $l_{max}$, of a hydrocarbon chain in a disordered structure such as a micelle is given approximately by

$$(l_{max} / \text{Å}) = 1.5 + 1.265 n_c \tag{5}$$

where $n_c$ is the number of carbon atoms. Thus, one expects a single layer of SDS molecules to be about 17 Å thick. A closely packed bilayer would extend to about 33 Å. It is clear that the observed layers, while thicker than a single molecule, are substantially thinner than a bilayer. A further interesting comparison can be made with the dimensions of micelles of SDS. Bergström and Pedersen [17] describe the micelles as oblate ellipsoids of revolution with the small half axis $a$ equal to 12 Å and the larger dimension $b$ of 23 Å for 1 %wt solutions in 0.1 M KBr. In this respect, the surface layer observed in the present experiments could correspond to micelles aligned on the surface. If such micelles were to flatten to disk-shapes and pack closely, the maximum packing would still be $\pi/4$ and it is interesting to note that this is not very different to the highest fraction of surfactant (0.77) found in the interfacial layer for the C-plane substrate at pH 3. More recently, it has been identified that scattering experiments alone are unable to distinguish unambiguously oblate and prolate ellipsoids with axial ratios in this range. [45] However, equivalent models would have similar anisotropy ratios and aggregation numbers, and in consequence at least one 'thin' dimension. As the present discussion is concerned with interfacial layers, the thermodynamic arguments that favor oblate ellipsoids and the other observations that suggest elongated micelles are present in solution when there are high concentrations of salt will not be reiterated here. Packing constraints are different for these alternatives but the interactions at the interface may modify the structure, the important conclusion to draw is that structures considerably thinner than twice the extended chain length predicted by Tanford are not uncommon. In SANS studies of micelles and in the measurements of the surface layer thickness described using neutron reflection, average dimensions are determined that include the dynamic fluctuations in both size and shape of aggregates. The similarity of the structures is also in agreement with the disorder in adsorbed layers described by Sperline et al. [10]

The changes in thickness shown in Figure 7b are relatively small compared to the reduction in the surface excess. This suggest that the change in the adsorbed amount is likely to arise from different lateral packing of aggregates rather than changes in the arrangement of the molecules perpendicular to the surface. The structure of adsorbed layers of surfactant at hydrophilic surfaces is sometimes discussed in terms of admicelles, hemimicelles or disordered bilayers. The layers observed both with and without salt are thin, and contain a significant fraction of water. The neutron reflection data, within error, is adequately modeled as a uniform disordered layer. The experiments do not resolve differences in the density profile for the heads, tails and counter ions of the surfactant.

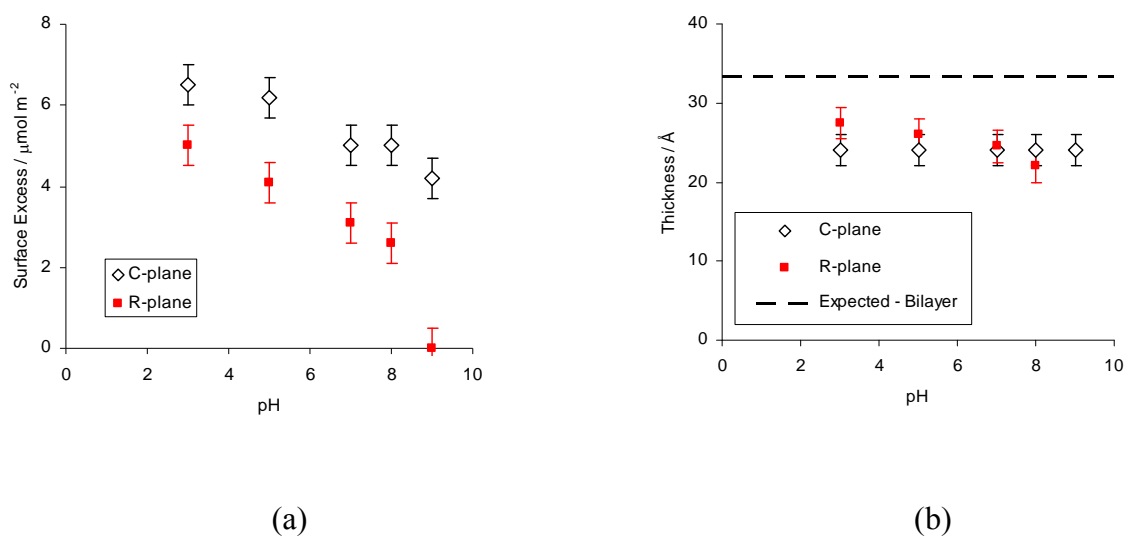

(a)                  (b)

Figure 7. Parameters from the model fits given in Tables II and III. (a) The variation of the adsorbed amount of SDS with pH of the solution is shown for the C and R plane interfaces. (b) The thickness of the adsorbed layers is plotted against solution pH and compared with the expected thickness of a bilayer of extended SDS molecules calculated according to Tanford. [44]

It is interesting to note that the change in adsorption with pH is quite different to that observed for another anionic surfactant, Aerosol-OT (sodium bis 2-ethylhexyl sulfosuccinate, AOT) at the sapphire solution interface. [37]  Only a small change in adsorbed amount was observed: the surface excess increased on making the solution either acidic or basic.  The changes for AOT with either added acid or base were similar to those seen with the same amount of extra salt ions and so were attributed to reduction in solubility of the surfactant.  Even without added salt, AOT displays a strong tendency to form dense well-ordered bilayers. [38]  In contrast, the adsorption of SDS is apparently controlled, in part, by the charge at the interface rather than being dominated entirely by the attractive hydrophobic interactions.  The gradual changes in surface excess with change of pH are similar to those observed in studies on colloidal alumina [7,8] that will usually have an amorphous surface layer.  This could imply that there are either a range of different chemical environments for surface OH that dissociate at different pH or that there is a dynamic regulation of the surface charge that is established with the surfactant ions.

The results for SDS without added salt indicate that there is a layer similar to that observed with salt although the coverage is less for the concentration measure well below the cmc.  The coverage and thickness are both slightly smaller but not different outside the experimental uncertainty.  There is an interesting study of SDS on the C-face of crystalline alumina with the surface force apparatus. [46]  This experiment measures the interaction between planar substrates submersed in solutions of surfactant.  The authors interpret their results as indicating bilayer adsorption with a thickness of 32 Å however the minimum separation that was observed was 20 Å and in the light of the present results might represent a thin bilayer or aggregate that bridges the two opposing sapphire substrates.

Lithium, sodium and cesium provide an interesting comparison as cations.  These were all measured at fixed ratios (0.3 and 1) to the cmc.  The heavier ions are expected to be less hydrated.  It is clear that the surface excess for CsDS is significantly higher than for the other surfactants and at the cmc there are almost no water molecules in the interfacial layer. The effect is more marked at the lower concentration of 0.3 cmc where the CsDS layer contains only 10% water. The results are broadly consistent with the picture from other studies [23,25] of less hydration and less dissociation of heavy counter ions.

In summary, the results presented show that SDS adsorbs to sapphire as a thin layer that is consistent with a structure of flattened micelles at the interface. The amount of adsorption depends on the pH of the solution and the observed changes in surface excess correspond with the expected surface charge. This correlation with surface charge extends to adsorption to different crystal faces of sapphire that have different isoelectric points.

**Acknowledgments**

We are grateful to the Institut Laue Langevin, Grenoble, France for allocation of beam time for the neutron reflection experiments and to Dr Giovanna Fragneto for assistance with the D17 reflectometer and Dr Max Wolff for help with the ADAM reflectometer.

**Tables**

Table I. Neutron scattering lengths of solvents, SDS and sapphire

| Material | Volume, $V$, / Å$^3$ | Scattering Length, $b$ / fm | Scattering Length Density, $\rho \times 10^{-6}$ Å$^{-2}$ |
|---|---|---|---|
| $H_2O$ | 30 | -1.7 | -0.56 |
| $D_2O$ | 30 | 19.1 | 6.35 |
| 50%$D_2O$ | 30 | 8.7 | 2.89 |
| A-$Al_2O_3$ | 42.5 | 24.3 | 5.71 |
| SDS | 474 | 15.9 | 0.34 |

*Scattering lengths are taken from Ref. [33].

Table II. Model fits for SDS adsorption on R-plane sapphire with 0.1M NaCl

| Solution | | $t$ / Å ± 2 | $\rho$ / $10^{-6}$ Å$^{-2}$ ± 0.1 | % Solvent ± 5% | $\Gamma$ / μmol m$^{-2}$ ± 0.5 | $A$ / Å$^2$ * ± 6 |
|---|---|---|---|---|---|---|
| pH=3 | $D_2O$ | 27 | 3.9 | 49 | 5.0 | 66 |
|  | 50%$D_2O$ | 28 | 1.8 | 46 |  |  |
| pH=5 | $D_2O$ | 25 | 4.2 | 54 | 4.1 | 81 |
|  | 50%$D_2O$ | 27 | 1.9 | 51 |  |  |
| pH=7 | $D_2O$ | 23 | 4.5 | 63 | 3.1 | 107 |
|  | 50%$D_2O$ | 26 | 2.2 | 64 | S |  |
| pH=8 | $D_2O$ | 22 | 4.6 | 66 | 2.6 | 128 |
| pH=9 | $D_2O$ | Not measurable | | | | |

*The area, $A$ refers to two molecules as a unit in the adsorbed layer.

Table III. Model fits for SDS adsorption on C-plane sapphire with 0.1M NaCl

| Solution | | $t$ / Å ± 2 | $\rho$ / $10^{-6}$ Å$^{-2}$ ± 0.1 | % Solvent ± 5% | $\Gamma$ / μmol m$^{-2}$ ± 0.5 | $A$ / Å$^2$ * ± 6 |
|---|---|---|---|---|---|---|
| pH=3 | $D_2O$ | 24 | 2.5 | 23 | 6.5 | 51 |
|  | 50%$D_2O$ | 24 | 1.3 |  |  |  |
| pH=5 | $D_2O$ | 24 | 2.7 | 27 | 6.2 | 54 |
| pH=7 | $D_2O$ | 24 | 3.4 | 41 | 5.0 | 66 |
| pH=8 | $D_2O$ | 24 | 3.4 | 41 | 5.0 | 66 |
| pH=9 | $D_2O$ | 24 | 3.9 | 50 | 4.2 | 79 |

*The area, $A$, refers to two molecules as a unit in the adsorbed layer.

**Table IV**. Model fits for LiDS, SDS and CsDS adsorption on C-plane sapphire with no added salt – measurements in $D_2O$

| Surfactant | Concentration / mM | $t$ / Å ± 2 | $\rho$ / $10^{-6}$ Å$^{-2}$ ± 0.1 | % Solvent ± 5% | $\Gamma$ / μmol m$^{-2}$ ± 0.5 | $A$ / Å$^2$ * ± 2 |
|---|---|---|---|---|---|---|
| LiDS | 2.6 | 24 | 5.2 | 80 | 2.1 | 160 |
|  | 8.8 | 24 | 1.4 | 17 | 8.6 | 39 |
| SDS | 2.4 | 20 | 3.4 | 51 | 4.4 | 77 |
|  | 8.2 | 20 | 2.2 | 31 | 6.0 | 56 |
| CsDS[#] | 2.1 | 26 | 1.1 | 10 | 10.0 | 33 |
|  | 6.9 | 24 | 0.4 | 0 | 10.4 | 32 |

*The area, $A$, refers to two molecules as a unit in the adsorbed layer. [#]Measured at 40 °C.

**Figure Captions**

Figure 1. Neutron reflection data for the clean R-plane substrate with a model fit that includes 5 Å roughness at the water interface. Data is shown for $H_2O$, 50% $D_2O$ and $D_2O$. The calculated fits are shown as solid lines.

Figure 2. Neutron reflection data for 1.4 mM SDS in 0.1 M NaCl/$D_2O$ adsorbed on R-plane sapphire with uniform layer fits. The model parameters for the fits, shown as solid lines, are listed in Table II.

Figure 3. Alternative representation of the data and fit for pH 3 from Figure 2 for the solution with 1.4 mM SDS and 0.1 M NaCl and the R-plane substrate. The plot of $RQ^4$ versus $Q$ allows the position of the peak and the layer thickness to be observed directly.

Figure 4. Neutron reflection data for 1.4 mM SDS in 0.1 M NaCl/$D_2O$ adsorbed on C-plane sapphire with uniform layer fits. The model parameters for the fits, shown as solid lines, are listed in Table III.

Figure 5. Neutron reflection data for SDS at concentrations of 2.4 mM (0.3 cmc) and 8.2 mM (1 cmc) in pure $D_2O$ adsorbed on C-plane sapphire with uniform layer fits. The model parameters for the fits are listed in Table IV.

Figure 6. Neutron reflectivity data for (a) LiDS and (b) CsDS in $D_2O$ measured at 25 and 40 °C respectively at concentrations of 0.3 cmc and 1 cmc.

Figure 7. Parameters from the model fits given in Tables II and III. (a) The variation of the adsorbed amount of SDS with pH of the solution is shown for the C and R plane interfaces. (b) The thickness of the adsorbed layers is plotted against solution pH and compared with the expected thickness of a bilayer of extended SDS molecules calculated according to Tanford. [44]